\documentclass{aa}  
\usepackage{amsmath}
\usepackage{graphicx}
\usepackage{txfonts}
\usepackage{natbib}
\usepackage{psfrag}
\newcommand{\kms}{\mbox{km\,s$^{-1}$}}

\newcommand{\Msun}{\mbox{M$_{\odot}$}}

\newcommand{\about}{\mbox{$\sim$}}

\def\HII        {\hbox{H\small{II}}}

\def\HCOP       {\hbox{HCO$^{+}$}}
\def\thirCO     {\hbox{$^{13}$CO}}
\def\water      {\hbox{H$_{2}$O}}
\def\source     {\hbox{G30.79\,FIR\,10}}           %%%% Rodrigo's command definitions  %%%%
\begin{document}

   \title{G30.79~FIR~10: A gravitationally bound infalling high-Mass star forming clump}

   \author{P.\,C.~Cortes
          \inst{1,3}
          \and
          R.~Parra\inst{2}
          \and 
          J.~R.~Cortes\inst{3}
          \and
          E.~Hardy\inst{3}
          }

   \institute{Departamento de Astronom\'ia y Astrof\'isica,
              Pontificia Universidad Cat\'olica de Chile,
              Casilla 306. Santiago 22. Chile.
              \email{pcortes@astro.puc.cl}
         \and
              European Southern Observatory, Alonso de Cordova 3107, 
              Vitacura, Casilla 19001, Santiago, Chile
         \and
              National Radio Astronomy Observatory, Joint ALMA Office, 
              Apoquindo 3846 piso 19, Las Condes, Santiago, Chile.\\
             }

% \abstract{}{}{}{}{} 
% 5 {} token are mandatory
 
  \abstract
  %context
  {The process of high-mass star formation is still shrouded in
    controversy. Models are still tentative and current observations
    are just beginning to probe the densest inner regions of giant
    molecular clouds. }
  % aims heading (mandatory)
  {The study of high-mass star formation requires the observation and
    analysis of high density gas. This can be achieved by the detection
    of emission from higher rotational transitions of molecules in
    the sub-millimeter. Here, we studied the high-mass clump G30.79 FIR
    10 by observing molecular emission in the 345\,GHz band. The goal is to
    understand the gravitational state of this clump, considering
    turbulence and magnetic fields, and to study the kinematics
    of dense gas. }
  % methods heading (mandatory)
  {We approached this region by mapping the spatial distribution of HCO$^{+}(J=4\rightarrow3)$,
    H$^{13}$CO$^{+}(J=4 \rightarrow 3)$, CS$(J=7 \rightarrow 6)$,
    $^{12}$CO$(J=3 \rightarrow 2)$, and $^{13}$CO$(J=3 \rightarrow 2)$
    molecular emission by using the ASTE telescope and by observing the
    $^{12}$C$^{18}$O$(J=3 \rightarrow 2)$, HCN$(J=4\rightarrow 3)$,
    and H$^{13}$CN$(J=4 \rightarrow 3)$ molecular transitions with
    the APEX telescope. }
  % results heading (mandatory)
  {Infalling
    motions were detected and modeled toward this source. A mean infall
    velocity of 0.5 \kms with an infall mass rate of $5\times10^{-3}$ \Msun/yr was
    obtained. 
    Also, a previously
    estimated value for the magnetic field strength in the plane of
    the sky was refined to be 855\,$\mu$G which we used to calculate a
    mass-to-magnetic flux ratio, $\lambda=1.9$, or super-critical. The
    virial mass from turbulent motions was also calculated finding
    M$_{\mathrm{vir}}=563$\,M$_{\sun}$, which gives a ratio of
    M$_{\mathrm{submm}}$/M$_{\mathrm{vir}}$=5.9. Both values strongly
    suggest that this clump must be in a state of gravitational collapse.
    Additionally, we estimated the
    HCO$^{+}$ abundance, obtaining $X$(HCO$^{+}$)= 2.4$\times 10^{-10}$.
  }
  % conclusions heading (optional), leave it empty if necessary 
   {}


   \titlerunning{G30.79 FIR 10 a bound clump?}
   \authorrunning{P. C. Cortes et al.}

   \maketitle
%
%________________________________________________________________

\section{Introduction}

It is well known that massive stars ($\gtrsim$10\,\Msun) form in giant
molecular clouds (GMCs), the largest molecular gas and dust complexes
in our galaxy.  The amount of gas involved in this process is several orders
of magnitude larger than in the low-mass star formation scenario,  
which significantly increases the uncertainties that makes its study difficult.  
In contrast to
the low-mass star formation case, where theory and observations have
established clear evolutionary steps \citep{McKee2007}, there are no
well defined stages for the evolution of proto-massive stars.
Presently, two different mechanisms are proposed as the dominant
process involved in the formation of massive stars. One mechanism is
the coalescence of small fragments
\citep{Bonnell1998,Stahler2000,Bonnell2004}. The other is accretion
directly onto a massive protostellar object, which is
supported by accumulating evidence for disks around proto-massive stars
\citep{Beuther2007,McKee2007}.

\noindent The rate of formation of massive stars can be up to several orders of
magnitude lower than low-mass stars, and they appear to be born in
clusters. The process is highly energetic and dynamic, a massive star will quickly perturb and
ionize its surrounding medium, which makes its study quite challenging
\citep{Wood1989,Hoare2007}.  Therefore, it is crucial to probe the
densest regions in high massive star forming clumps to understand
their physical and chemical conditions. In this regard, high
rotational levels from molecules emitting in the sub-millimeter
windows are the ideal tracers to go for. The combination of optically
thick and optically thin isotopomers of the same molecular species can be used to infer
physical properties from star formation sites.

\noindent The magnetic field is likely the least known physical parameter
in star formation. While its presence seems to be ubiquitous within the
ISM, with strengths ranging from a few $\mu$G to mG
\citep{Crutcher1999a}, there are surprisingly few observations.
Moreover, while it is still unclear what role is played by
the magnetic field in the formation of low-mass stars, the degree of
uncertainty is even higher for high-mass star formation owing to the
lack of observations. However, we can say with certainty that 
magnetic fields have been observed towards high-mass star forming regions
\citep{Cortes2006a,Girart2009}. Therefore, it is
important to incorporate it as a relevant physical parameter in 
observations and models of high-mass start formation.

\noindent In this paper we present a multi-line study of the
high-mass star forming region G30.79 FIR 10
The aim of this paper is to understand the 
gravitational state of this clump as well as to study the
kinematics of the dense gas. We also use previous inteferometric
observations of polarized dust emission to include information
about the magnetic field toward this region. 
The paper is organized as follows, Sect. 1 is the
introduction, Sect. 2 presents the source, Sect. 3 the
observational procedure. In Sect. 4 we present the results, while in
Sect. 5 we discuss the abundance of HCO$^{+}$. The kinematics of the gas is 
studied in Sect. 5, where we present evidence for
infall motions and discuss the likelihood of outflows. Additionally,
we refined a previous estimation of the magnetic field strength in the plane of
the sky for this source \citep{Cortes2006a}, and use it to evaluate the
gravitational state of this clump. Finally, Sect. 6
presents the summary and conclusions.

%__________________________________________________________________

\section{The source}

G30.79 FIR 10 (hereafter G30.79) is a massive molecular complex
located within the W\,43 region. It involves an \HII\ region-molecular
cloud complex near $l=31^{\circ}, b=0^{\circ}$, with several 
far-infrared sources, of which G30.79 is the most massive and densest
component.  Figure~\ref{1} presents an overview of our observations
superposed over the 350\,$\mu$m continuum map from \citet{Motte2003}.
\citet{Liszt1995} observed G30.79 in \HCOP\ and \thirCO\, concluding
that the presence of several rings and shells in the dense molecular
gas was a disturbance product of star formation. \citet{Vallee2000}
observed the dust continuum emission in this source at 760\,$\mu$m
using JCMT. They found linear polarization of about 1.9\% with a
position angle (P.A.) of 160$^{\circ}$. \citet{Mooney1995}
observed this source at 1.3\,mm using the IRAM 30\,m telescope,
detecting a total flux density of 13.6\,Jy; their wide field map shows
the clump and the extended \HII\ region in the G30.79 complex.  \water\
masers have been observed toward this region \citep{Cesaroni1988}, which
are within a half arc-second of the peak in the \citet{Mooney1995} map.
Additionally, \citet{Ellingsen2007} detected 6.7\,GHz methanol masers
in the proximity of G30.79, which is considered to be a
signature for massive star formation. No centimeter radio-continuum
emission seems to be associated with FIR\,10, suggesting that the
source could be in an early stage of evolution. However, the maser emission
already indicates that star formation has started and outflows may be
present. \citet{Motte2003} mapped the
W43 main complex in dust continuum emission at 1.3\,mm and 350\,$\mu$m
with the IRAM 30\,m and CSO telescopes, respectively. They also mapped the
HCO$^{+}$ $J=3\rightarrow2$ line and measured H$^{13}$CO$^{+}$
$J=3\rightarrow2$ towards prominent dust maxima.  One of the maxima,
W43-MM1, corresponds to G30.79  and is the compact fragment we
observed with ASTE.  \citet{Motte2003} found $V_{\mathrm{lsr}}=
98.8$\,km\,s$^{-1}$, $\Delta v=5.9$\,km\,s$^{-1}$ (from
H$^{13}$CO$^{+}$), T$_{\textnormal{dust}} \sim 19$\,K, M$\sim
3600$\,M$_{\sun}$, and $n(\textnormal H_{2}) \sim 8.8 \times
10^{6}$\,cm$^{-3}$.  They estimated the virial mass to be M$_{\rm
  vir}$\about 1000\,\Msun, suggesting that this compact fragment
should be in a state of gravitational collapse unless there are other
sources of support in addition to kinetic energy.  \citet{Cortes2006a}
mapped G30.79  in dust polarized emission at 1.3\,mm using the
BIMA interferometer founding a polarization pattern, which suggests an
hour-glass morphology for the field. They also estimated the magnetic
field strength in the plane of the sky to be \about1.7\,mG, which gave
a statistically corrected mass-to-magnetic flux ratio of 0.9 or
critical, where by critical we mean the equilibrium value between 
self-gravitation and magnetic field support for the cloud.

\section{Observations}

\subsection{ASTE observations}
G30.79 FIR 10 was observed during September 2006 using the Atacama Sub-millimeter Telescope
Experiment (ASTE) from the National
Astronomical Observatory of Japan (NAOJ)
\citep{Kohno2005}.  The telescope is located at {\em Pampa la bola} in
the Chilean Andes plateau reserve for Astronomical research at 4900
meters of altitude.  ASTE is a 10\,m diameter antenna equipped with a
345\,GHz double side band SIS-mixer receiver. We 
simultaneously observed $^{12}$CO$(J=3\rightarrow2)$ and
HCO$^{+}(J=4\rightarrow3)$, CS$(J=7\rightarrow6)$ and
$^{13}$CO$(J=3\rightarrow2)$, and H$^{13}$CO$^{+}(J=4\rightarrow3)$ with
a beam size of $\sim 22^{\prime \prime}$, and a velocity resolution
of 0.1\,\kms, setting the MAC (which is a XF-type digital spectro-correlator) 
to a bandwidth of 128 MHz.  The pointing accuracy was
in the order of 2$^{\prime \prime}$, with Uranus used as the pointing
source. The observations were done by performing raster maps (position 
switching) with a
grid spacing of $15^{\prime \prime}$ (see
Table\,\ref{obspar} for map sizes), operating the telescope remotely
from the ASTE base in San Pedro de Atacama under good weather
conditions  (precipitable water vapor or PWV $<$ 1\,mm). Our reference position was
$(\alpha,\delta)=(18^{\mathrm{h}}47^{\mathrm{m}}46.9^{\mathrm{s}},
-1^{\circ}54^{\prime}29.1^{\prime \prime})$ (J2000), which coincides
with the peak dust emission reported by
\citet{Mooney1995,Motte2003,Cortes2006a}. Unfortunately, the off position used
to subtract the continuum had emission in the $^{12}$CO$(J=3\rightarrow2)$ line.
However, only the $^{12}$CO line-wings  were affected (see
discussion below).  The M17SW molecular complex
was used as an intensity calibrator. All temperatures are presented as
T$_{\mathrm{mb}}=$T$^{*}_{\mathrm{A}}/\eta_{\mathrm{mb}}$, where
$\eta_{\mathrm{mb}}=0.71 \pm 0.07$. Initial data reduction and
calibration was done using the NEWSTAR package, and the calibrated data were
later exported into our own software for analysis and plotting.

\subsection{APEX observations}
Observations were performed during the first week of August 2008
using the Swedish Heterodyne Facility Instrument (SHFI) mounted on the
Atacama Pathfinder Experiment telescope (APEX) \citep{Gusten2006b}, located
at llano de Chajnantor in the Chilean Andes. We tuned SHFI at 329.3\,GHz in
order to detect the $^{12}$C$^{18}$O$(J=3 \rightarrow
2)$ molecular transition, to 354.5 GHz for HCN$(J=4\rightarrow3)$, and
to 345.3 for H$^{13}$CN$(J=4\rightarrow3)$. The spectrometer was set up
to 8192 channels with a resolution of 0.1\,km\,s$^{-1}$.  The main
beam efficiency is $\eta$=0.73$\pm$0.07 as measured by the APEX staff,
with a pointing accuracy better than 2$^{\prime \prime}$ and a
beam size of $19^{\prime \prime}$.  Jupiter and R-Aql were used as intensity
calibrators where the observations were calibrated by the usual
chopper-wheel method. The observations were done in raster mode with
spacings of $15^{\prime \prime}$ from the same reference position used
for the ASTE observations. The initial data reduction was done with
the GILDAS-CLASS reduction package and the final analysis with our own
software tools.
%%
%% Figure 1
%%

\begin{figure*}
\centering
\includegraphics[width=0.45\hsize]{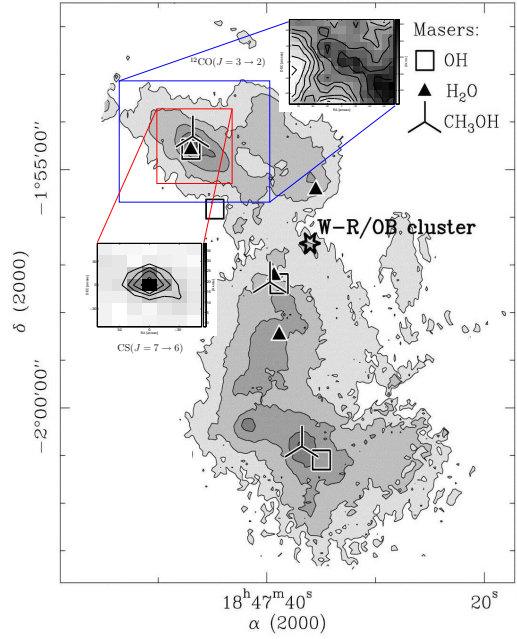}
\par\psfrag{CS76}[m][c][0.5]{CS$(J=7\rightarrow 6)$}
\par\psfrag{12CO32}[m][c][0.5]{$^{12}$CO$(J=3\rightarrow 2)$}
\caption{Overview of the \source\ massive star forming region. The
  main panel represents the 350 $\mu$m continuum emission from
  \citet{Motte2003}, Figure~1. Contours and grayscale are 12, 24, 48, 96, and 200 
  Jy\,beam$^{-1}$. The cluster of massive stars is indicated by the star
  symbol, while masers are indicated by the symbols presented at the upper
  right of the figure. Superimposed are our new $^{12}$CO$(J=3\rightarrow
  2)$ and CS$(J=7\rightarrow 6)$ integrated velocity maps to show the
  extent of our observations. 
  }
\label{1}
\end{figure*}

%
%Table 1
%
%\include{tab1}
\begin{table*}
  {\centering
  \caption{Line parameters}
  \label{obspar}
  \begin{tabular}{c c c c c c c c c}
    \hline  \hline
    Line             & Transition          & Frequency   & Observed Area               & T$_{\mathrm{peak}}$ & $\int$ T $dv$ & V$_{\mathrm{peak}}$ & $\Delta V$ \\     % table heading
    &            &   [GHz]   &  [arcsec$^2$] & [K] & [K \kms\ ] & [\kms\ ] & [\kms\ ] \\
    \hline                         % inserts single horizontal line
    HCO$^{+}$        & $(J=4\rightarrow3)$ & 356.7342880 & $200^{\prime \prime} \times 150^{\prime \prime}$ & 5.7 $\pm 0.1$  & 42.6 $\pm 1.8$ & 94.8 & 8.4\\   % inserting body of the table
    H$^{13}$CO$^{+}$ & $(J=4\rightarrow3)$ & 346.9983381 & $90^{\prime \prime} \times 90^{\prime \prime}$ & 1.2 $\pm 0.02$ & 5.5  $\pm 0.6$ & 98.1 & 3.0\\
    HCN              & $(J=4\rightarrow3)$ & 354.5054779 & $200^{\prime \prime} \times 150^{\prime \prime}$ & 3.8 $\pm 0.1$ & 73.5 $\pm 13.8$ & 93.8 & 9.5     \\   % inserting body of the table
    H$^{13}$CN       & $(J=4\rightarrow3)$ & 345.3397750 & $30^{\prime \prime} \times 30^{\prime \prime}$ & 1.4 $\pm 0.1$ & 14.8  $\pm 0.4$ & 99.1 & 4.8 \\
    CS               & $(J=7\rightarrow6)$ & 342.8828503 & $90^{\prime \prime} \times 90^{\prime \prime}$ & 3.6 $\pm 0.2$ & 38.7 $\pm 0.9$ & 98.5 & 3.8\\
    $^{12}$C$^{16}$O & $(J=3\rightarrow2)$ & 345.7959899 & $200^{\prime \prime} \times 150^{\prime \prime}$ & 16 $\pm 2.2$& 317 $\pm 21.1$ & 98.8 & 9.6 \\
    $^{13}$C$^{16}$O & $(J=3\rightarrow2)$ & 330.5879652 & $90^{\prime \prime} \times 90^{\prime \prime}$ & 8.5 $\pm 1.2$ & 125 $\pm 7.6$  & 97.4 & 6.3 \\
    $^{12}$C$^{18}$O & $(J=3\rightarrow2)$ & 329.3305525 & $30^{\prime \prime} \times 30^{\prime \prime}$ & 3.1 $\pm 0.2$ & 42.9 $\pm 10.0$    & 98.8 & 4.7 \\
    \hline
  \end{tabular}
  }\\ \\
  {Parameters for all molecular line observations. 
    The values for T$_{\mathrm{peak}}$ and $\Delta V$ are
    calculated from Gaussian fits  to the central pointing
    or $(18^{\mathrm{h}}47^{\mathrm{m}}46.9^{\mathrm{s}},
-1^{\circ}54^{\prime}29.1^{\prime \prime})$. For HCN and HCO$^{+}$,
   the V$_{\mathrm{peak}}$ was taken from
   Gaussian fits done to the most intense component, while the line-width corresponds to
   the whole line. For $^{12}$CO and 
   $^{13}$CO the fit was done over the whole spectrum.}
\end{table*}

\section{Results}

\subsection{$^{12}$CO, $^{13}$CO, and $^{12}$C$^{18}$O $(J=3
  \rightarrow 2)$ results}

\begin{figure*}
  \centering
  \includegraphics[width=0.45\hsize]{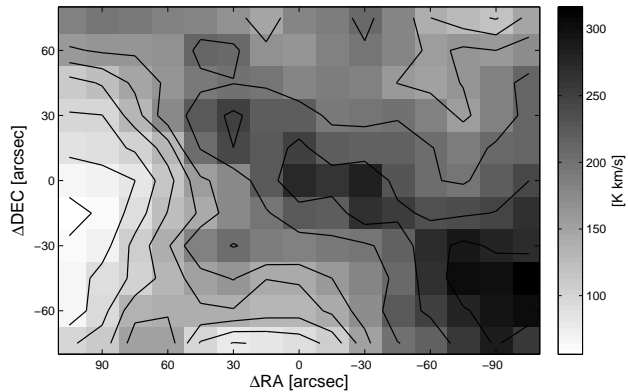}
  \caption{Velocity integrated $^{12}$CO$(J=3 \rightarrow 2)$ map. The
    axis units are offsets in arc-seconds from the reference
    position $(18^{\mathrm{h}}47^{\mathrm{m}}46.9^{\mathrm{s}},
-1^{\circ}54^{\prime}29.1^{\prime \prime})$ in J2000 coordinates. The gray-scale is in K\,\kms, while the contours are 3,
    4, 5, 6, 7, 8, 10, 12, and 14$\sigma$ with $\sigma=20$ K\,\kms.}
  \label{2}
\end{figure*}

\begin{figure*}
  \centering
  \includegraphics[width=0.45\hsize]{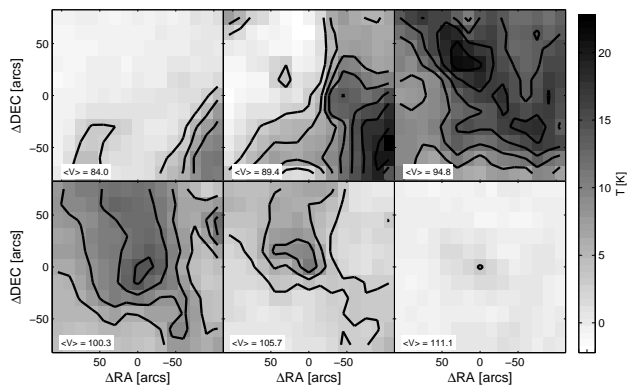}
  \caption{Channel maps for the $^{12}$CO line emission. Maps are
    shown as averages of 50 channels giving a velocity resolution of
    5.4\,\kms for each map. The average velocity in each plot is shown
    by the box in lower left corner of each map. Both $\Delta \alpha$
    and $\Delta \delta$ are arc-second offset with respect to the reference
    position. The pixel scale is in K degrees and is normalized to the
    maximum and minimum values from all the maps set. The contours are
    from 3, 6, 9, 12, and 15$\sigma$, where $\sigma=0.9$\,K. for all
    maps and calculated from a velocity range with no emission.  }
\label{3}
\end{figure*}

The $^{12}$CO $(J=3 \rightarrow 2)$ line was detected all over
the sampled region, covering an area of
$3.5^{\prime}\times2.5^{\prime}$ around the peak from the dust
emission at 1.3\,mm (see Fig.~1) The velocity integrated emission
map is presented in Fig.~\ref{2} where the $^{12}$CO emission was
integrated over the complete available velocity range (43 to 154\,\kms\
). The peak in the map corresponds to 317\,K\,\kms\ located at about
$85^{\prime \prime}$ to the south-west from the reference position, while the peak
at the reference position is 234\,K\,\kms.  Overall we see most of
the emission clustered in a south-west to north-east elongated
pattern, which is consistent with the morphology of the dust emission
seen in the maps from \citet{Motte2003} and
\citet{Mooney1995}. However, the peak in our $^{12}$CO map does not
coincide with the peak in dust emission (located at our reference
position). 
Figure~\ref{3} shows velocity channel maps binned every
5\,\kms\ (or 50 channels) with a resolution of 5.4\,\kms\ for each
map with respect to the original resolution of 0.1\,\kms.  The channel maps
are presented in increasing velocity from 84 to 106\,\kms\, covering
the most intense features in the $^{12}$CO emission.  By examining
Fig.~\ref{3}, a velocity gradient from the south-west to north-east
can be seen with each successive channel where the peaks in emission are
at 89 and 95\,\kms\ and at almost opposite spatial
locations. Particularly interesting is the intense emission seen at
the south-west part of the map. According to \citet[see Fig.~3 and
also our Fig. \ref{1}]{Motte2003}, the giant \HII\ region produced
by a Wolf-Rayet cluster of massive stars has  not yet reached the G30.79 
clump. However, the \citet{Mooney1995} maps put the \HII\ region at an
interface with the clump.  Because we could not sample beyond
105$^{\prime \prime}$ west, we can only speculate about the nature of
this intense $^{12}$CO$(J=3\rightarrow2)$ emission  and whether it is related to an interface between this clump and the  H II region without further observations.

%%
%% Figure 2
%%

%%
%% Figure 3
%%

\begin{figure}
\centering
\includegraphics[width=0.9\hsize]{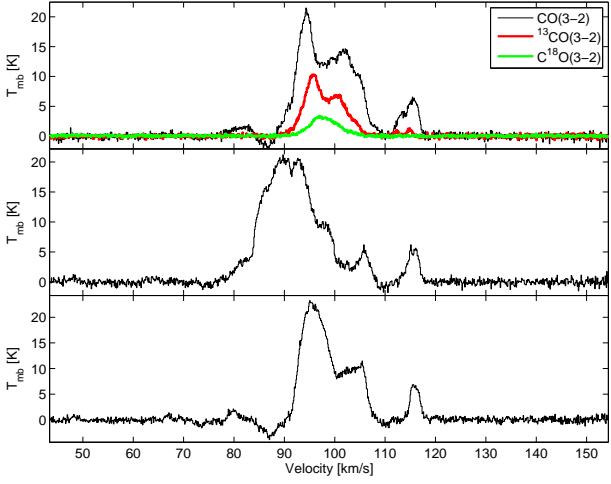}
\caption{Spectra from $^{12}$CO $(J=3 \rightarrow 2)$ are
  presented. Each spectrum corresponds to an average of nine pointings
  centered at offsets, from upper to bottom, ($\Delta \alpha$, $\Delta
  \delta$)=(0,0), (-90,-40), and (30,40), where the offsets are in
  arc-seconds. The velocity resolution for all the spectra is taken to be
  0.1 \kms. The first panel also shows the spectra from $^{13}$CO
  $(J=3 \rightarrow 2)$ in red and C$^{18}$O $(J=3 \rightarrow 2)$ in
  green. Clearly visible is the negative dip feature, likely due to emission
  in the off position, seen in both the upper
  and lower panels for the $^{12}$CO $(J=3 \rightarrow 2)$  spectra.
  The component at 115 km s$^{-1}$ is also seen in both $^{12}$CO
  and $^{13}$CO but not in C$^{18}$O.  }
\label{4}
\end{figure}

To study the velocity structure at selected locations in the
integrated emission map, we present $^{12}$CO spectra
from three positions in the map in Fig. \ref{4}. These positions correspond to
averages over nine pointings centered at
$(\Delta\alpha,\Delta\delta)=(0,0)$, $(-90,-40)$ and $(30,40)$
arcsec. The (0,0) offset plot corresponds to the spectrum
over the reference position and had superposed $^{13}$CO and
$^{12}$C$^{18}$O spectra as well.  The emission is strong with a peak
of 16\,K for the $^{12}$CO, 8.5\,K for $^{13}$CO, and 3.1\,K for
$^{12}$C$^{18}$O, obtained by fitting Gaussian profiles. The
remaining spectra correspond only to $^{12}$CO and are taken from the
maxima seen at 89 and 95 km\,s$^{-1}$ velocity channel maps.
We did not obtain data from the other CO isotopomers
at these positions.  
We immediately appreciate that the $^{12}$CO line
profiles are complex and have many components with the emission appearing to
be optically thick.  But owing to the large angular extension of the
whole W43 region, we did not find a suitable off position in our
observations. This is clearly seen in the dip features 
at the line wings of our spectra (see Fig. \ref{4} upper panel). 
While the emission at the off position is weak compared to the peak
of the line, it is strong enough to disrupt the line-wings.
In principle this does not affect the
integrated intensity map, but it will prevent us from concluding about
possible outflow motions from the CO line-wings. We also see that all 
spectra show blue-shifted peaks relative to $V_{\mathrm{lsr}=98.8}$ \kms. The
associated peak velocities are 94.24\,\kms\ for (0,0), 89.69\,\kms\
for (-90,-40), and 95.00\,\kms\ for (30,40). 
All spectra show broad
line-wings, particularly at offset (-90,-40), and a high velocity
component at $v$=115\,\kms. This peculiar emission may indicate
outflow motions, but as mentioned before, it is difficult to 
be decisive due to contamination from the off position.

In mapping $^{13}$CO $(J=3 \rightarrow 2)$, we were only able to cover
an area of $1.5^{\prime} \times 1.5^{\prime}$ around the reference
position.  The maximum is found at 124.7 K\,\kms\ located at
30$^{\prime \prime}$ west from the center. The emission also seems to
follow a south-west to north-east orientation like $^{12}$CO do. Yet,
we cannot confirm the correlation due the lack of equal coverage for
our $^{13}$CO.  Fig.~\ref{4}, first panel, shows a spectrum
corresponding to the nine central pointings, or 30$^{\prime \prime}
\times 30^{\prime \prime}$, in $^{12}$C$^{18}$O and $^{13}$CO overlaid
on the $^{12}$CO spectrum.  The same complex features seen in
$^{12}$CO are seen in the $^{13}$CO line profile, particularly the
high velocity component at 115\,\kms. Signs of self-absorption near
the center of the line seem to be present, but not the strong
absorption seen in the $^{12}$CO profile at the edges of the line
wings. From these features, it is likely that the $^{13}$CO emission
that we detected towards the center is also optically thick.  In
contrast, $^{12}$C$^{18}$O is single-peaked, with an almost Gaussian 
profile, showing no trace of
self-absorption, which suggests a likely optically thin emission.

An additional feature of the CO data is a consistent blue-shifted peak
emission. It has been suggested that optically thick lines with a
blue-shifted profile may indicate infall motions
\citep{Leung1977}. Both $^{12}$CO and $^{13}$CO emission are similar
in complexity showing the same gradient in velocity from south-west to
north-east, broad line wings, and blue asymmetry in the line profile
($V_{\mathrm{lsr}}=$98.8\,\kms). Particularly interesting is that
the line profiles in addition to their blue-shifted peak are both 
self-absorbed,
which is also often seen in infall candidates \citep{Klaassen2007}. 
%On the
%other hand, optically thin lines, as our $^{12}$C$^{18}$O observations
%might be, with
%a blue-shifted peak are also seen when infalling is present.
Considering that our offset (0,0) corresponds to the peak in dust
emission at both 1.3\,mm and 350\,$\mu$m, which covers most of the
continuum spectrum from pre-stellar cores, it is likely that the condensation
we are seeing is actively forming stars. Unfortunately, we cannot
resolve the number of fragments due to the coarse $22^{\prime \prime}$
beam of the ASTE telescope. Our APEX observations, with their
$19^{\prime \prime}$ resolution will not help either because previous
interferometric, 4$^{\prime \prime}$ resolution, observations also did
not resolve the core \citep{Cortes2006a}. However and due to the
expected multiplicity in high-mass star forming cores, it is unlikely
that we are seeing only a single object.

%%
%% Figure 4
%%

%%
%%Figure 5
%%

\begin{figure*}
\centering
\includegraphics[width=0.45\hsize]{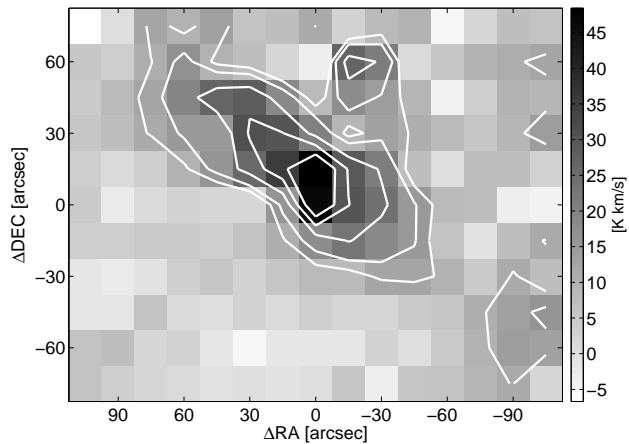}
\caption{Velocity integrated emission map for HCO$^{+}(J=4 \rightarrow
  3)$ over the whole velocity range (45 to 150\,\kms) is shown.  The
  pixel scale is in units of K\,\kms, while the contours
  are on levels of 3, 4, 6, 8, and 10 $\times$ $\sigma$, with
  $\sigma=3.7$\,K\,\kms.  }
\label{5}
\end{figure*}

\subsection{HCO$^{+}(J=4 \rightarrow3)$ and
  H$^{13}$CO$^{+}(J=4\rightarrow3)$ results}

\begin{figure}
\centering
\includegraphics[width=0.9\hsize]{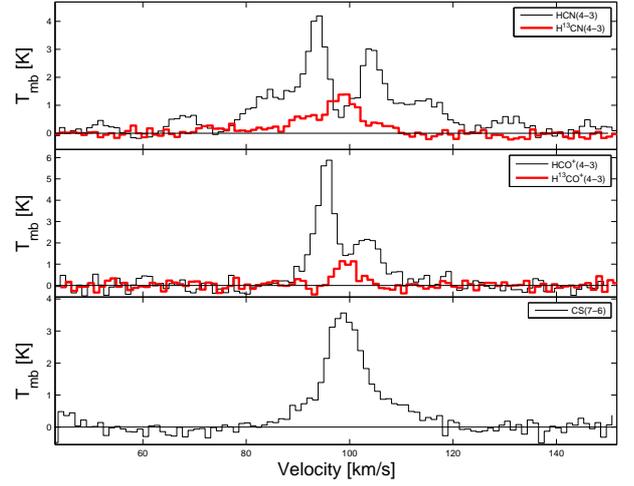}
\caption{
High density molecular gas spectra from the G30.79 FIR 10 clump.
The {\em upper} panel presents the HCN$(J=4 \rightarrow
  3)$ spectrum. In
  thick red lines the H$^{13}$CN$(J=4 \rightarrow 3)$ spectrum is
  superposed. 
  The {\em central} panel presents the HCO$^{+}(J=4 \rightarrow
  3)$ spectrum. In
  thick red lines the H$^{13}$CO$^{+}(J=4 \rightarrow 3)$ spectrum is
  superposed. The peak brightness
  temperature of H$^{13}$CO$^{+}$ is 1.2\,K. The {\em lower} panel
  presents the CS$(J=7 \rightarrow 6)$ spectrum. All spectra were
  taken at the reference position. Both HCO$^{+}$ and the HCN spectra were 
  binned every five channels, giving a velocity resolution of 0.54 \kms, while
  CS, H$^{13}$CO$^{+}$ and H$^{13}$CN were binned every ten channels, giving a
  velocity resolution of 1.1 \kms.
  }
\label{6}
\end{figure}

We observed HCO$^{+}(J=4 \rightarrow 3$) simultaneously with $^{12}$CO
$(J=3 \rightarrow 2)$, which also  covered  an area of
$3.5^{\prime}\times2.5^{\prime}$.  The velocity integrated emission
map is presented in Fig.~\ref{5}. The morphology of the emission
follows the same orientation south-west to north-east as the $^{12}$CO
observations, but without the extension of $^{12}$CO.  Clearly, the
strongest emission is clustered at the center of the map, to the
north-east, while some hint of emission over $3\sigma$ seems to appear
at the south-west, roughly at ($\Delta \alpha,\Delta \delta$)=(-80$^{\prime
  \prime}$,-40$^{\prime \prime}$), where the strongest $^{12}$CO
emission is located. The map morphology is also consistent with the
HCO$^{+}(J=3\rightarrow2)$ map of \citet[see Fig.~2]{Motte2003}.
Figure~\ref{6}, central panel, shows the spectrum from the
reference position for both HCO$^{+}$ and H$^{13}$CO$^{+}$. 
The line profile is clearly non-Gaussian, showing
evident self-absorption. It is likely that the HCO$^{+}$ emission is
optically thick.  No hint of emission is seen at the off position, which is same
off position used for $^{12}$CO. 
Superposed in thick lines is the H$^{13}$CO$^{+}
(J=4 \rightarrow 3)$ spectrum with a peak brightness temperature of
1.2\,K at V=98.1\,\kms.
The H$^{13}$CO$^{+}$ emission is only significant (over the 3$\sigma$ level)
at the reference position. The velocity associated with the peak can
be considered to be at the $V_{\mathrm{lsr}}$ within the boundaries of
the bin. The emission is also coincident with the strong absorption dip
in HCO$^{+}$, suggesting self-absorption. Additionally, because
the line profile appears to be Gaussian in shape, it
is likely that the H$^{13}$CO$^{+}$ emission is optically thin.

%%
%%Figure 6
%%

\subsection{CS$(J=7 \rightarrow 6)$ results }
\begin{figure*}
\centering
\includegraphics[width=0.45\hsize]{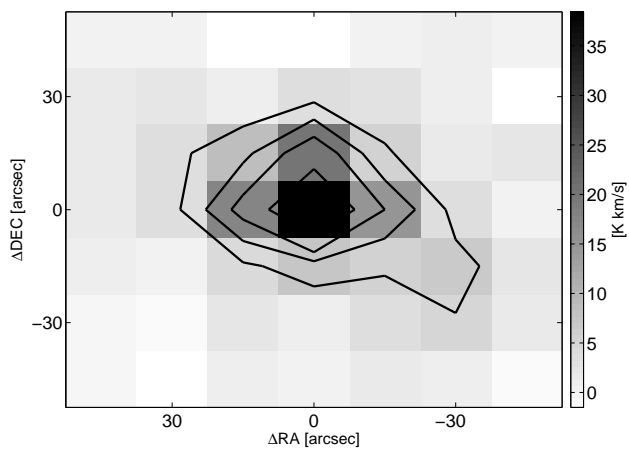}
\caption{Velocity integrated emission map for CS$(J=7 \rightarrow 6)$
  over the whole velocity range (45 to 150\,\kms). The pixel map has units
of K \kms. The contour levels are 3, 6, 9, and 15 $\times \sigma$, with 
$\sigma=1.7$ K \kms.  }
\label{7}
\end{figure*}

The integrated velocity map for CS$(J=7 \rightarrow 6)$ is presented
in Fig.~\ref{7}. As for H$^{13}$CO$^{+}$, the CS
emission is fairly compact, arising only from the center of the map, which
suggests that most of the activity is coincident with the peak of
the dust emission.  The spectra from the
central 30$^{\prime \prime} \times 30^{\prime \prime}$ are shown in Fig. \ref{6} in
the lower panel.  
The CS emission presents  a peak brightness temperature of 3.6\,K at
V=98.1\,\kms with some excess emission in its line-wings,
which may be due to  outflowing motions. 
Note that the CS emission is not self-absorbed as seen
with HCO$^{+}$ and HCN. { It is possible that the molecule is not 
abundant enough to become self-absorbed. However, another possibility is
that the CS gas is bound inside the dense core, which would explain why its 
emission is not widespread (as indicated by our map).}

\subsection{HCN$(J=4 \rightarrow3)$ and
  H$^{13}$CN$(J=4\rightarrow3)$ results}

\begin{figure*}
\centering
\includegraphics[width=0.5\hsize]{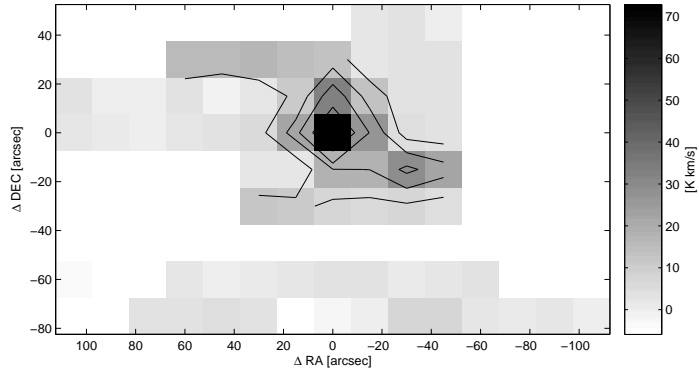}
\caption{Velocity integrated emission map for HCN$(J=4 \rightarrow 3)$.
The integration is done over 60 to 140 \kms to include all the broad
HCN emission.  The pixel map has units
of K \kms. The contour levels are 3, 6, 9, and 15 $\times \sigma$, with
$\sigma=3.0$ K \kms. The color scale indicates the emission level with 
white used for positions not observed or with zero emission. }
\label{hcn}
\end{figure*}

The HCN$(J=4 \rightarrow3)$ was observed in position-switching mode over and area of
$200^{\prime \prime} \times 150^{\prime \prime}$ to sample
the same region as HCO$^{+}$. We looked for emission only in
the most significant areas within the region.
The H$^{13}$CN$(J=4\rightarrow3)$
was only observed over the central $30^{\prime \prime} \times 30^{\prime \prime}$,
where we expected to find most of the emission.
Figure \ref{hcn} presents the velocity integrated emission map
for HCN. While the emission is mostly concentrated at the center, there is a hint
for a gradient along the south-west to north east direction as with HCO$^{+}$ and CO.
The extent of the emission appears to be midway between the HCO$^{+}$$(J=4 \rightarrow 3)$
 and the CS$(J=7 \rightarrow 6)$ where the CS map presents the most compact
morphology.
Fig. \ref{6} shows in its upper panel the HCN$(J=4 \rightarrow3)$
overlaid by the H$^{13}$CN$(J=4\rightarrow3)$ spectrum in red. The HCN emission
is clearly opticaly thick with a self-absorption feature as indicated
by the optically thin H$^{13}$CN. The double peaked HCN
spectrum has a stronger blue peak at 93.8 \kms; while
the H$^{13}$CN is peaked at 98.3 \kms, both taken from Gaussian fits.
This type of spectrum is often seen toward infall candidates
like the  well-studied source B335 \citep{Zhou1993}.

\section{Analysis and discussion}

\subsection{The HCO$^{+}$ abundance}

In the current paradigm, it is suggested that carbon bearing molecules
will freeze onto grain surfaces, forming ices, in cold and dense cores
\citep[see e.g.][]{Bergin2007}; while some other molecular abundances,
like HCO$^{+}$, might get enhanced by non-thermal motions such as
outflows and/or infall \citep{Rawlings2004}. In order to investigate
the abundance of HCO$^{+}$ toward this region, we estimate the
physical parameters associated with this molecule. The optical depth for
HCO$^{+}$ is estimated by the ratio between the optically thick and
optically thin brightness temperatures by using \citep{Choi1993}

\begin{equation}
  \label{od}
  \frac{T_{\mathrm{mb}4\rightarrow3}}{T_{\mathrm{mb}(13)4\rightarrow3}}=
  \frac{1 - e^{-\tau_{43}}}{1 - e^{-\tau_{43}/X}},
\end{equation}

\noindent where $T_{\mathrm{mb}4\rightarrow3}$ is the brightness
temperature from the HCO$^{+}$ line,
$T_{\mathrm{mb}(13)4\rightarrow3}$ is the H$^{13}$CO$^{+}$ line
temperature, and $X$ is the ratio between $X$=[$^{12}$C]/[$^{13}$C]
taken to be 50, a value representative for the molecular ring at the 4 kpc inner
galaxy \citep{Wilson1994}. 
Because the HCO$^{+}$ profile is self-absorbed,
there is a significant uncertainty in deriving the line intensity. Two possible approaches
can be followed: either we pick the strongest peak, 
or a Gaussian profile can be adjusted to the
whole line by masking out the self absorption dip. \citet{Purcell2006} found
no major differences from these two approaches in their results for a sample of
40 high-mass star forming regions. Thus, we used the strongest peak for the HCO$^{+}$
line intensity in this calculation. From 
these numbers we obtained $\tau_{\mathrm{hco^{+}}}^{43}=11.3$
 indicating an optically thick  HCO$^{+}$ line, which is to be
expected due to the self-absorption seen in the line profile (see Fig. \ref{6}).  
Also, note that both 
the HCO$^{+}$ and the H$^{13}$CO$^{+}$ line were obtained from the
reference position and with a beam size of about $18^{\prime \prime}$, which encloses
 the dust core detected from the interferometric observations
of this region \citep[see Figure 1]{Cortes2006a}.

\noindent To obtain the column density for the upper rotational level, we use
\citep{Goldsmith1999}

\begin{equation}
  N_{4}=\frac{8\pi k \nu^{2}}{hc^3 A_{43}} \left(\frac{\Delta \Omega_a}{\Delta \Omega_s}
  \right) \left(\frac{\tau_{43}}{1 - e^{-\tau_{43}}} \right) \int T_{\mathrm{mb}} dv,
\end{equation}

\noindent where $\nu$ is the line frequency,
$A_{43}$ is the Einstein spontaneous emission coefficient 
($3.6269\times10^{-3}$\,s$^{-1}$ for HCO$^{+}(J=4 \rightarrow 3)$),
$\Delta \Omega_a$ is the antenna solid angle and $\Delta \Omega_s$ is the
source solid angle for the core which gives a beam filling
factor of $0.73$ for HCO$^{+}$.  The total column density is then given by

\begin{equation}
  \label{cdensity}
  N=\frac{N_{u}Z}{g e^{-E_u/T}},
\end{equation}

\noindent where $Z$ is the partition function, $g=2J + 1$ is the
statistical weight of the upper level, $E_u=42.8$ K is the energy
for the upper level \citep{Schoier2005}, and $T=30$ K is the dust
temperature used by \citet{Cortes2006a} to calculate the total
hydrogen column density $N$(H$_2$). This is justified under the LTE
assumption. For linear molecules, the partition function is well
approximated by $Z=kT/hB_0$, where $B_0=45$\,GHz is the rotational
constant for HCO$^{+}$ taken from \citet{Defrees1982}. In this way, we
estimated a HCO$^{+}$ column density $N$(HCO$^{+}$) = 1.5 $\times
10^{14}$ cm$^{-2}$.  
From a sample of 40 high-mass star forming
regions selected from methanol maser detections, \citet{Purcell2006}
estimated the HCO$^{+}$ column densities ranging from 2.5$\times 10^{14}$
cm$^{-2}$ to 71 $\times 10^{14}$ cm$^{-2}$.  Our column density estimation
is consistent with the smallest values obtained in their work. 
Now, by using the column density of hydrogen
derived from dust emission or $N$(H$_2$) =
6.5 $\times 10^{23}$ cm$^{-2}$, we calculate a total abundance for
HCO$^{+}$ or $X$(HCO$^{+}$) = 2.4 $\times 10^{-10}$.  As previously
mentioned, it is expected that carbon bearing molecules will get
depleted toward cold and dense regions. We can quantify this by
calculating the depletion factor $f_D$ defined as the ratio between
the {\em average} fractional abundance and the observed
one. \citet{Lucas1996} and \citet{Liszt2000} determined a
$X_{\mathrm{ISM}}=2 \times 10^{-9}$ from a sampling of
HCO$^{+}(J=1\rightarrow0)$ toward the diffuse ISM in absorption
against background QSOs.  By using this value for $X_{\mathrm{ISM}}$,
we find $f_D=8.4$, which suggests that a moderate depletion of HCO$^{+}$
toward this clump when compared to other high-mass star forming regions
such as G305 with $f_D=1.5$ \citep{Walsh2006,Bergin1997}.
However,  we maybe underestimating
the HCO$^{+}$ column density due to the uncertainty introduced by the brightness 
temperature used from the observed self absorbed profile, which would lead a lower
depletion value for HCO$^{+}$. 
{ These results, along with the lack of infrared
sources detected at this clump, and with no compelling evidence for outflow emission, 
even though maser emission has been observed (see 
next section), suggest an early (pre-hot core) evolutionary
stage for G30.79 FIR 10}. In an advanced stage, it is expected that the
radiation coming from the central condensation will evaporate the ice
mantles increasing the abundance of molecules like HCO$^{+}$, creating what is known
as a hot core. Note though, that the detection of
maser emission toward this clump suggests that the star formation process has
 already started. 

\subsection{An outflow in G30.79 FIR 10?}

One of the signatures of high-mass star formation are the powerful
molecular outflows observed toward these regions
\citep{Shepherd2007,Bourke1997}. Because FIR~10 is a massive core, it
is likely that outflowing motions are present or will develop over time.
However, to untangle the outflow motions from our molecular emission
observations is certainly not trivial.  The
choice of sub-millimeter emission lines as a tool to study this region
allow us to separate the most dense components from the rest of the
molecular core. 
Particularly, $^{12}$CO $(J=3\rightarrow2)$ has been
successfully used to trace outflowing motion from star forming regions
\citep{Choi1993}. However, our $^{12}$CO $(J=3\rightarrow2)$ results are
inconclusive because  of emission at  the 
off position, which  affected the 
line wings in our spectra. Also, the
high velocity component seen at V=115\,\kms, which is \about20\,\kms\ over
the $V_{\mathrm{lsr}}$ and is only seen in $^{12}$CO and $^{13}$CO
but not observed either in HCO$^{+}$, HCN, or CS. It is
likely that this component is not dense enough to excite either of these
lines; indeed, this high velocity component might be optically
thin. Although it could be interpreted as high velocity outflow
emission, its widespread spatial distribution makes this
interpretation un-likely. Another possibility is that this component
corresponds to a foreground object, or cloud, which is not associated with
this clump. Additionally, it is difficult to accelerate the gas to
such high velocities without dissociation.

\begin{figure*}
  \centering
  \includegraphics[width=0.45\hsize]{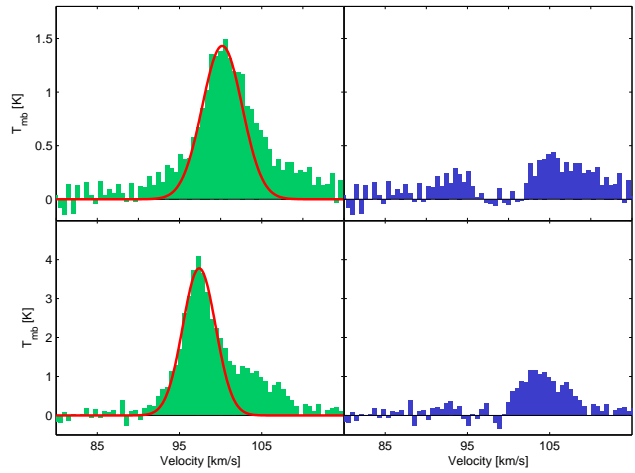}
  \caption{Residual emission after Gaussian substraction is shown in 
both panels. The {\em lower panel} shows the HCO$^{+}(J=4 \rightarrow
    3)$ profile with an adjusted Gaussian to the left. To the right,
    in the same panel, the residual emission from the Gaussian
    subtraction is presented. The {\em upper panel} shows CS$(J=7
    \rightarrow 6)$ profile with the adjusted Gaussian profile and
    its residual emission to the right.  The $\sigma$ value is
    calculated to be 0.08\,K. Both spectra were taken as averages over
    an area of $30^{\prime \prime} \times 30^{\prime \prime}$. }
  \label{9}
\end{figure*}

Outflows are discovered through their signature in the line-wings of
molecular emission lines. Fig.  \ref{9} shows the HCO$^{+}$ and CS 
spectra with their corresponding Gaussian fittings and the residual
emission in their line-wings. Between 100 and 110 \kms we
found some significant emission over the $3\sigma$ level, with
$\sigma=0.08$ K, in both CS and HCO$^{+}$ spectrum. However,
the blue-shifted part of the emission, for velocities lower
than 95 \kms, does not show significant traces of residual emission.
Even though the red-shifted excess emission may be due to non-thermal
motions, the bipolar nature of an outflow must be clearly stated,
which we cannot do with these data.
Therefore, we cannot yet come to a conclusion about the presence of
a bipolar outflow towards this source. Our lack of spatial resolution
due to the distance to G30.79 FIR 10 might be the reason behind this.
However, outflows are ubiquitous 
in the high-mass star forming regions, so it is likely that they
are present or will develop over time. 
Additional observations of
shocked excited chemistry, such as SiO, SO, or SO$_{2}$ may help confirming
an outflow in this clump.

%%
%% Figure 8
%%

\subsection{Infall motions}

\begin{figure*}
  \centering
  \includegraphics[width=0.65\hsize]{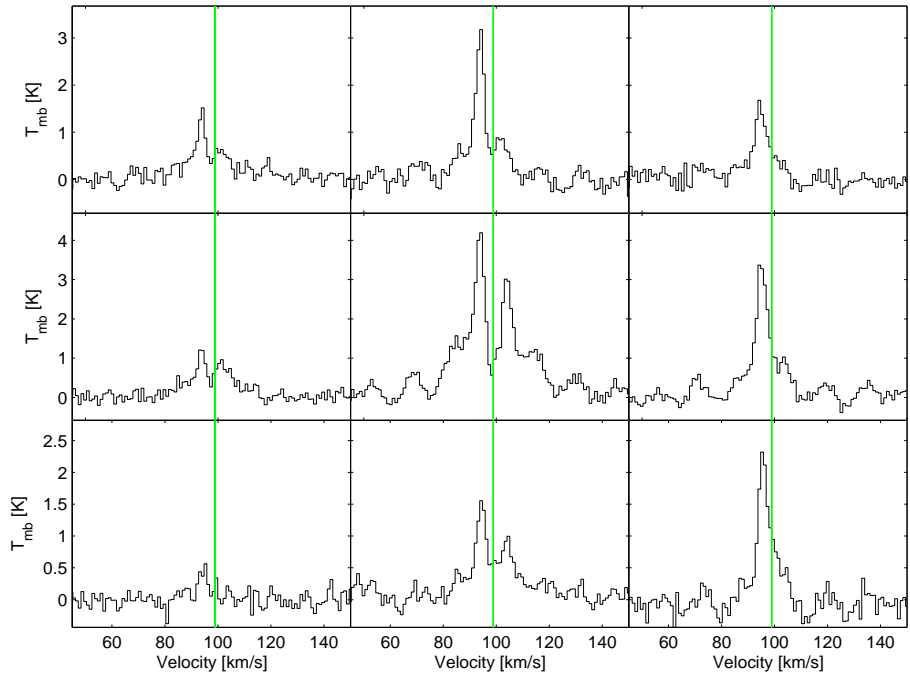}
  \caption{The central $48^{\prime \prime} \times 48^{\prime \prime}$ region
compose of 9 pointings separated by $15^{\prime \prime}$ with a beam size
of $18^{\prime \prime}$ each is shown. The green line represents the systemic
velocity at 98.8 \kms; while the scale is given in K (main beam temperature)}
  \label{hcnpanel}
\end{figure*}

It is not clear whether high-mass stars
forms through accretion or through a different process such as
coalescence of less massive fragments. This situation is difficult to
distinguish due to the physical complexities involved in the evolution of
high density gas and dust. The study of the kinematics and dynamics at
the earliest phases, along with the detection of accretion disks,
could clarify this uncertainty.  In this scope, the determination of
infalling motions is a first step toward the identification of
collapsing pre-stellar objects. The characterization of these motions
is a challenging topic, the current avenue towards investigating infalling is the
study of asymmetries in molecular line profiles.  \citet{Leung1977}
suggested that an asymmetry in the line profile toward the blue may
indicate the presence of infalling motions.
Thus, the low excitation, red-shifted
infalling layers of gas in the front part of the cloud absorbs
some of the emission from the rest of the gas. This red-shifted
self-absorption is what makes the spectrum present a brighter blue peak.
To quantify this asymmetry, the calculation of the normalized
line velocity difference has been used by many authors
 \citep{Fuller2005,Szymczak2007,Wu2007}.

\begin{equation}
\label{dv}
\Delta V_{\mathrm{be}}=\frac{V^{\mathrm{max}}_{\mathrm{thick}} - V^{\mathrm{max}}_{\mathrm{thin}}}
{\Delta V_{\mathrm{thin}}},
\end{equation}

\noindent where $V^{\mathrm{max}}_{\mathrm{thick}}$ is the velocity from the
peak emission of the optically thick line,
$V^{\mathrm{max}}_{\mathrm{thin}}$ is the velocity from the peak
emission of the optically thin line, and $\Delta V_{\mathrm{thin}}$ is
the FWHM line width from the optically thin molecular tracer. A
negative $\Delta V_{\mathrm{be}}$ will correspond to blue-shifted gas
velocities, which are indicative of infalling motions.  
We calculated $\Delta V_{\mathrm{be}}$ for the high density
tracers HCN, CS, and HCO$^{+}$ using H$^{13}$CO$^{+}$ and H$^{13}$CN 
as the optically thin tracers. These molecules have high dipole moments
and therefore require higher critical densities to get excited. 
In particular we used H$^{13}$CO$^{+}$
as the thin component for HCO$^{+}$ and H$^{13}$CN for CS and HCN to 
keep ion-molecules and neutral molecules separate.
\noindent Table \ref{sk} gives the values of $\Delta V_{\mathrm{be}}$ for all
three lines. 
The difficulty in
separating infalling evidence from the line asymmetry arises because
other dynamic phenomena such as rotation and outflows will
also produce red and blue asymmetries in the molecular profiles. 
 However, infalling is the only motion which will produce
only blue asymmetries. 
The expected infall signature will manifest in spectral
lines with a double peaked profile where the blue peak is stronger
than the red peak and the dip is due to self absorption. 
A blue asymmetry is seen in all spectra, with the highest
 value in HCO$^{+}$, but
the characteristic infalling double peak profile is only seen in both HCN and HCO$^{+}$.
Both appear to be self absorbed and have a stronger blue component 
(see Fig. \ref{6}).
The case of HCO$^{+}$ is interesting,
the red component is almost non-existent having the
largest blue asymmetry in all three molecules. The CS spectrum is single peaked to
the blue, but does not show
the self-absorbed features. It has the lowest intensity, which could
be cause by carbon and sulfur depletion in this object. 
In this way, it may not have enough abundance to self-absorb.
\citet{Evans2003}  suggested a procedure to evaluate whether infall is likely
or not in a star forming core. We followed the procedure checklist and found
that our HCN observations, taken at the reference position, 
meet the conditions required for infall.
Moreover, the same properties are observed in the HCN spectra
from the surrounding pointings (see Fig. \ref{hcnpanel}). 
Almost all of them present double peaked profiles with a stronger blue peak, 
clearly suggesting  infalling. 
Together with  the previous checklist 
we can conclude that our HCN spectra meet the requirements to suggest infall movements
toward this source. 
Thus, we quantified the infall parameters through modeling. To do this,
we applied the simple infalling model devised by \citet{Myers1996} and later improved
by \citet{DiFrancesco2001}. In this model,
the clump is approximated by two infalling gas layers, a front and a rear layer,
 with a central source modeling the pre-stellar core. 
Thus, the observed brightness temperature is quantified by the following equation, where 
the subscripts $``$f", $``$r", $``$C", and $``$cmb" stand for the front 
layer, the rear layer, the central source, and the cosmic background.

\begin{table}
\label{param}      % is used to refer this table in the text
\caption{Parameters from the model fit}              % title of Table
{\centering
\begin{tabular}{c c c c c c}        % centered columns (4 columns)\hline\hline                 % inserts double horizontal lines
\hline\hline
$J(T_{\mathrm{f}})$ & $J(T_{\mathrm{r}})$ &$v_{\mathrm{f}}$ & $v_{\mathrm{r}}$ & $\tau_{0}$ & $\sigma$\\
$[\mathrm{K}]$ & [K] & [\kms] & [\kms] & - & [\kms] \\
\hline                        % inserts single horizontal line
   24.6 & 17.5 & 0.7 & -0.3 & 2.0 & 3.4  \\
\hline                                    %inserts single line
\end{tabular}
} \\ \\
{The parameters from the best fit obtained by adjusting the \citet{DiFrancesco2001} infalling model to our HCN($J = 4 \rightarrow 3$) spectrum.}
\end{table}

\begin{equation}
\label{model}
\Delta T_{\mathrm{B}} = (J_{\mathrm{f}} - J_{\mathrm{C}})[1 - e^{-\tau_{\mathrm{f}}}] +
(1-\Phi)(J_{\mathrm{r}} - J_{\mathrm{cmb}})[1 - e^{-(\tau_{\mathrm{r}} + \tau_{\mathrm{f}})}].
\end{equation}

\begin{figure*}
  \centering
  \includegraphics[width=0.35\hsize]{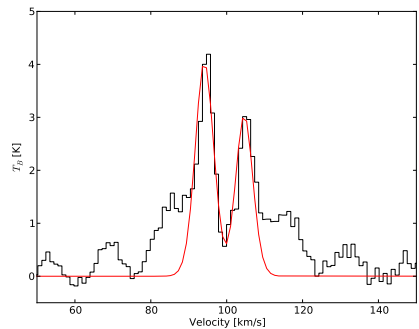}
  \caption{Central HCN spectrum from G30.79 FIR 10. Overlaid in thick lines is the best fit for the
           infalling model.  }
\label{modelfit}
\end{figure*}

\noindent The main terms in this equation are
  the Planck excitation temperature given by $J_{\mathrm{i}} = 
T_{0}/[\exp{(T_{0}/T_{\mathrm{i}})} - 1]$ with $T_{0} = h\nu/k$, and $T_{\mathrm{i}}$ corresponding to 
either $T_{\mathrm{f}}$, $T_{\mathrm{r}}$, $T_{\mathrm{c}}$, and $T_{\mathrm{cmb}}$. Also,
$J_{\mathrm{C}} = \Phi J_{\mathrm{c}} + (1 - \Phi)J_{\mathrm{r}}$ where $\Phi$ is the beam
filling factor of the continuum source, which due to our large beam size is assumed to be 0. 
The $\tau_{\mathrm{i}}$ expressions correspond to the optical depths, which
we assumed to be Gaussian \citep[following][]{Myers1996}.
Thus, the front and rear optical depths are given by

\begin{eqnarray}
\tau_{\mathrm{f}} = \tau_{0} \exp{\left[ \frac{-(v - v_{\mathrm{f}} - v_{\mathrm{lsr}})^2}
{2\sigma^2} \right]} \\
\tau_{\mathrm{r}} = \tau_{0} \exp{\left[ \frac{-(v + v_{\mathrm{r}} - v_{\mathrm{lsr}})^2}
{2\sigma^2} \right]},
\end{eqnarray}

\noindent where $\tau_{0}$ is the peak optical depth for both
the front and the rear layers, $v_{\mathrm{f}}$ and  $v_{\mathrm{r}}$ 
are the infalling velocity for both slabs and $\sigma$ is the velocity dispersion.
The model was fitted in a two step process, first by doing a
multi Gaussian fit component, which provided the input parameters to later adjust Eqn. \ref{model} 
to our data by minimizing
the $\chi^{2}$ function through the Levenburg-Marquardt algorithm \citep{Press2002}. 
The best fit to the HCN spectrum is shown in Fig. \ref{modelfit}, and the fit parameters
are presented in Table \ref{param}.
Note that this is the simplest possible model that we can use to 
characterize infalling gas. Moreover,
the structure of the HCN line is complex and, even though the features seen in the 
line wings may come from  outflowing motions, we did not consider them in 
this simple model.

\noindent The infalling velocity for both slabs is estimated by the infalling model
to be $v_{\mathrm{f}}=0.7$ \kms and $v_{\mathrm{r}}=-0.3$ \kms.
A simple estimation can also be done by using the analytical model
derived by \citet{Myers1996} for a contracting cloud.
By using Eqn. 9, from their model, we
estimated an infalling velocity of 0.5 \kms. This result is in good agreement with 
the the values obtained by adjusting the model to our data. 
The next step is to estimate the mass infall rate, which we do by 
following \citep{Klaassen2007}

\begin{equation}
\label{massinfallrate}
\dot{M}=\frac{\mathrm{d}M}{\mathrm{d}t} \approx \frac{M}{t}=\frac{\rho V 
v_{\mathrm{in}}}{R}= \frac{4}{3}\pi n_{\mathrm{H}_2} \mu m_{\mathrm{H}} R^2 v_{\mathrm{in}},
\end{equation}

\noindent where $\mu=2.35$ is the mean molecular weight, $R=0.3$ pc 
is the geometric radius taken from the interferometric observations, and 
$n_{\mathrm{H}_2}$ is the gas density. The infall rate obtained for this 
source is about $5\times10^{-3}$ \Msun /yr. Both infall velocity and infall mass rate are consistent
with the values seen at other high-mass star forming regions, like G10.47,
which has an 
estimated infall velocity of  1.8  \kms and an infall rate
of $10^{-2}$ \Msun /yr \citep{Klaassen2007}. 
\citet{Barnes2010}
estimated infall rates of $3 \times 10^{-2}$ \Msun/yr with an infalling
velocity of 0.3 \kms for the G286.21+0.17 molecular clump.
Both regions present similar gas densities as G30.79 FIR 10 ($1.9 \times
10^5$ cm$^{-3}$ for G286.21+0.17 and $7.2 \times 10^5$ cm$^{-3}$ for
G10.47) and similar mass estimations ($4.1 \times 10^4$ \Msun for G286.21+0.17
and $7.3 \times 10^{3}$ \Msun\ for G10.47  \citealp[as estimated by][]{Shirley2003}). 
These results are consistent with our findings where these regions appear
to be in a similarly evolutionary stage.

\begin{table}
  \caption{Normalized velocity difference.}
\label{sk}      % is used to refer this table in the text
{\centering                           % used for centering table
\begin{tabular}{c c}        % centered columns (4 columns)
\hline\hline                 % inserts double horizontal lines
Line & $\Delta V_{\mathrm{be}}$ \\     % table heading
\hline                        % inserts single horizontal line
   HCN$(J=4\rightarrow3)$              & -1.1  \\   % inserting body of the table
   HCO$^{+}(J=4\rightarrow3)$          &  -1.2 \\   % inserting body of the table
   CS$(J=7\rightarrow6)$               & -0.1  \\
\hline                                    %inserts single line
\end{tabular}
}\\ \\
  {The asymmetry of line profiles is presented by
    the calculation of $\Delta V_{\mathrm{be}}$, as shown in Eqn. \ref{dv},
    where the optically thin
    species correspond to H$^{13}$CO$^{+}$ for HCO$^{+}$ and 
    H$^{13}$CN for HCN and CS. The values used to compute the blue excess
    are taken from Gaussian fits to the corresponding spectra.
    }              % title of Table
\end{table}

\subsection{Dynamical state of G30.79 }

\subsubsection{The magnetic field}

The dynamical state of high-mass star forming cores is difficult
to establish. The accumulation of a large amount of molecular gas
is still poorly understood, as is what triggers the gravitational 
collapse. Moreover, it is known that magnetic fields are present in these
regions with strengths substantially larger than in low-mass star forming
regions \citep{Lai2001,Cortes2006a}. Thus, the magnetic field can have
a profound effect on the dynamical evolution of the gas as postulated
by ambipolar diffusion theories \citep{Ciolek1993}. Our object of study
is not an exception to the above. Magnetic fields have been observed 
through polarized emission from dust towards this core 
\citep{Vallee2000,Cortes2006a}. However, it is not yet  clear how important the
field is with respect to turbulence in the high-mass star forming process, but
its presence alone is a reason to consider its effects.

From the interferometric observations of polarized dust emission,
\citet{Cortes2006a} found a smooth polarimetric pattern centered at
the G30.79 FIR 10 clump. Assuming magnetic alignment of dust grains, they proposed
an hourglass morphology for the field. Additionally, they
estimated the magnetic field strength on the plane
of the sky using the Chandrasekhar-Fermi method \citep{Chandrasekhar1953}, 
finding a value of $B_\mathrm{pos}=1.7$\,mG. In this way,
the mass-to-magnetic flux ratio was calculated to be $\lambda = 0.9$
or critical. In this work, we refined the previously described
estimation by considering our molecular line observations. Before
stating this improvement, we will briefly introduce the 
Chandrasekhar-Fermi method by the following equation.

\begin{equation}
\label{chf}
B_{\mathrm{pos}}=9.3\frac{\sqrt{n(\mathrm{H_{2}})}\Delta V}{\delta \phi},
\end{equation}

\noindent where $B_\mathrm{pos}$ is the magnetic field strength in
$\mu$G, $n$(H$_2$) is the volumetric gas number density in cm$^{-3}$,
$\Delta V$ is the FWHM velocity width in km s$^{-1}$, and $\delta
\phi$ is the polarization position angle dispersion in
degrees. \citet{Cortes2006a} used $n$(H$_2$) = 4.5$\times
10^{5}$ cm$^{-3}$ calculated from their dust emission, while $\delta
\phi=21^{\circ}.9$ was obtained from the polarization data. The
line-width, $\Delta V$, was taken from the
H$^{13}$CO$^{+}(J=3\rightarrow 2)$ observations by \citet{Motte2003}
for the G30.79 clump (called MM1 and listed at Table~2 in their work).
It is here where we improve the estimation by using $\Delta V$
derived from our H$^{13}$CO$^{+}(J=4\rightarrow 3)$ observations.
The $(J=4\rightarrow 3)$ transition traces higher density molecular
gas, while being better coupled to the field than H$^{13}$CN or other neutral 
species. Also, the H$^{13}$CO$^{+}(J=4\rightarrow 3)$ emission is optically 
thin, which might not be the case for the $(J=3\rightarrow 2)$ transition.
Additionally, this higher transition comes from a smaller 
beam size, which engulfs only the FIR 10 clump (of about 20$^{\prime \prime}$
in size), while the $(J=3\rightarrow 2)$ transition comes a higher
beam size ($\sim 30^{\prime \prime}$),
 which might be including velocity components not associated
to the field perturbations. Thus, by using $\Delta V=3.0$\,\kms\
we obtain a value for $B_{\mathrm{pos}}=$855\,$\mu$G, which gives a
statistically corrected mass-to-magnetic flux of $\lambda=1.9$, or
super-critical, consistent with the infalling scenario
suggested in this work. Thus, the magnetic field is not
strong enough to support the core.

\begin{figure*}
  \centering
  \includegraphics[width=0.45\hsize]{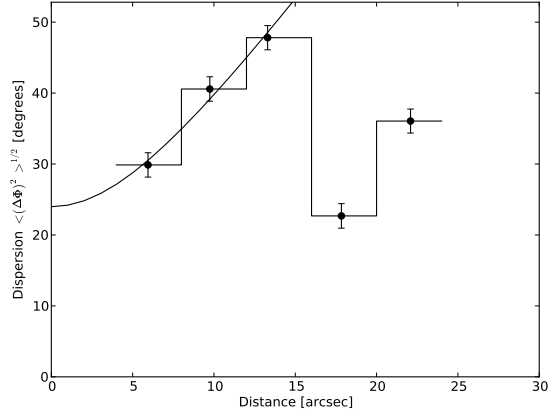}
  \caption{The polarization position angle dispersion function
is shown here in closed circles along with its respective error bars.
The dispersion function points are join with thick lines, while superposed
is the fit for Equation \ref{hildebran} to the first 3 points.}
  \label{bturb}
\end{figure*}

While the polarization pattern observed by \citet{Cortes2006a} is smooth,
it is unclear how distorted by turbulence the field might be. 
%Even though their
%interferometric observations provide the highest angular resolution yet, the angular
%scale is still coarse and the effect of turbulence maybe smooth-out by the beam.
This source
is 5.5 kpc away which gives 0.03 pc/1$^{\prime \prime}$ or 0.1 pc/beamsize.
This length scale is larger than the predicted turbulence length scale of
1 mpc \citep{Lazarian2004,Li2008}. Therefore, the dispersion in the
field lines, or the turbulent component of the field, cannot be directly
seen  from the polarization pattern.
\citet{Hildebrand2009} developed an approximation to calculate the plane 
of the sky dispersion function
for the magnetic field in turbulent molecular clouds. By using polarized
emission from dust, they calculated a dispersion function for the polarization
position angle as a quadratic function of
the length scale. Two main assumptions are made in this calculation. The length scale is larger
than the turbulence, or correlation length scale, and smaller than the
length scale associated with large scale variations of the field.
Our BIMA polarization data satisfy both assumptions, the interferometric
observations have a length scale on the order of 0.1 pc while the field varies
over 1 pc distances (as seen from the polarization pattern). 
This approximation is stated in Eqn. 2 in \citet{Hildebrand2009}.
From this approximation, the ratio between the turbulent component of the
field on the plane of the sky and the main magnetic field strength can be 
estimated as

\begin{equation}
\frac{\left< B_{t}^2 \right>}{B_0} = \frac{b}{\sqrt{2 - b^2}},
\label{hildebran}
\end{equation}

\noindent where $b$ is the turbulent contribution to the dispersion function,
$\left< B_{t}^2 \right>$ is the turbulent component of the field, and 
$B_0$ is the main magnetic field. 
We applied this method to our BIMA data by calculating the dispersion function
for polarization position angle as shown in Fig. \ref{bturb}. The value of
$b=0.42$ is given by the intersect of the best fit to the data points, and gives
a turbulent-to-main magnetic field ratio of $\left< B_{t}^2 \right>/B_0=0.3$.
This result suggest that 30\% of the total magnetic field strength is in
the turbulence.

\subsubsection{The gravitational equilibrium in G30.79 FIR 10}

This source is one of the few where information about the magnetic field and
turbulence is present. Thus, we should be able to characterize with confidence
the state of gravitational equilibrium in this source. In order to quantify this,
we will follow the analysis as stated by \citet{Mckee1992} and \citet{Bertoldi1992}:

\begin{equation}
  2(\mathcal{T} - \mathcal{T}_0) + \mathcal{M} + W = 0
\end{equation}

\noindent where $\mathcal{T}$ is the total kinetic energy including
thermal and non-thermal motions, $\mathcal{T}_0$ correspond to the
surface term for the kinetic energy, or external pressure,
$\mathcal{M}$ is the magnetic energy, and $W$ is the gravitational
energy term. We will follow \citet{Motte2003} and assume the giant \HII\
region present in the W43 complex does not reach the clump.
In this way, we can drop $\mathcal{T}_0$
out the equation. Only the kinetic and magnetic terms
remains to balance gravity and as was said before, we have information
about both of them.

\citet{Motte2003} estimated the virial mass without knowing the
magnetic field, stating that G30.79 FIR 10 is likely to be bound. To estimate
the virial mass they used $M_{\mathrm{vir}}=5R\sigma^2/G$ from
\citet{Bertoldi1992}, which we refined by deriving our own values for
$R=0.3$\,pc (obtained from  the interferometric map
of dust emission from \citealt{Cortes2006a} as half the size of
the continuum core or $10^{\prime \prime}$), 
and $\sigma=1274$ m s$^{-1}$ from our
H$^{13}$CO$^{+}$ observations ($\sigma$ is not the FWHM value), obtaining
$M_{\mathrm{vir}}=563$ M$_{\sun}$. Comparing this with the mass of the
clump taken to be M$_{\mathrm{submm}}$ = 3300\,M$_{\sun}$
\citep{Cortes2006a} we obtained the ratio of the virial mass to the
dust mass to be M$_{\mathrm{submm}}$/M$_{\mathrm{vir}}=5.9$. Values of
M$_{\mathrm{submm}}$/M$_{\mathrm{vir}} > 0.5$ are considered to remain
gravitationally bound according to \citet{Pound1993}. At the same time,
we recall our previous calculation of the mass-to-magnetic flux
ratio, which was found to be super-critical.  This means that there is
not enough energy in the turbulence and/or  in the magnetic field to
balance gravity. Therefore and even though the uncertainties involved
in the total mass estimations for this object, the results
suggest that G30.79 FIR 10 is bound and must be undergoing
gravitational collapse. The whole picture appears to be self-consistent,
we have detected infalling motions into the clump where neither turbulence or
the magnetic field have enough energy to counterbalance gravity.
%An
% additional interesting point is that the
%turbulent term is of similar significance than the magnetic term
%suggesting that both turbulence and magnetic energy are in
%equipartition.
%                                     Two column figure* (place early!)
%______________________________________________ Gamma_1 (lg rho, lg e)
\section{Summary and conclusions}

We mapped the G30.79 FIR 10  molecular clump embedded in the W43
mini-starburst by mapping the molecular emission from $^{12}$CO$(J=3
\rightarrow 2)$, $^{13}$CO$(J=3 \rightarrow 2)$, C$^{18}$O$(J=3
\rightarrow 2)$, CS$(J=7 \rightarrow 6)$, HCO$^{+}(J=4\rightarrow3)$,
H$^{13}$CO$^{+}(J=4\rightarrow3)$, HCN$(J=4\rightarrow3)$, and 
H$^{13}$CN$(J=4\rightarrow3)$.  Emission from $^{12}$CO is
extended presenting a gradient
from south-west to north-east, consistent with the dust morphology of
the core. Even though we mapped $^{13}$CO from a smaller region
surrounding the center, its emission appears to follow $^{12}$CO, which is
also optically thick. The C$^{18}$O emission was also mapped at the
central $30^{\prime \prime} \times 30^{\prime \prime}$. Its line
profile is almost Gaussian without any of the complex features seen in
the previous CO isotopomers and is likely optically thin. We found a high
velocity component at 115\,\kms, which was also present all over our sampled
region. We interpreted this emission as a foreground cloudlet not
associated with our clump. The emission from HCO$^{+}$ is 
optically thick, with an optical depth $\tau=11.3$, but more compact
than $^{12}$CO. This is expected due to the higher densities required
to excite this molecule.  Additionally, both CS and H$^{13}$CO$^{+}$
are even more compact, with the emission coming mostly from the central
$30^{\prime \prime} \times 30^{\prime \prime}$ and consistent with the
position of the dust maxima obtained by others. By using our
observations, we estimated the HCO$^{+}$ abundance $X$(HCO$^{+}$)=2.4 
$\times$ $10^{-10}$, a depletion factor $f_D=8.4$,
which is consistent with estimations done toward other high-mass star
forming regions \citep{Purcell2006}. 

Outflows were looked for in the line-wings of the observed molecular
lines toward the center of G30.79 FIR 10. All HCO$^{+}$, HCN, and CS
profiles present excess emission in their line wings. However, the 
bipolar nature of this emission is inconclusive at the $3\sigma$ level.
Infalling motions were also looked for by studying the profile asymmetry
of our molecular observations. The blue asymmetry was estimated by calculating
the normalized velocity difference (see Table \ref{sk}). The clearest evidence for infall is
given by the HCN spectra. Toward the center and surroundings pointings,
the emission is double peaked with the blue peak
stronger than the red peak (see Fig. 6 and \ref{hcnpanel}). 
By using the \citet{Myers1996} improved model by \citet{DiFrancesco2001},
we estimated an infall velocity of 0.5 \kms with an infall rate of 
$5\times10^{-3}$ \Msun /yr.

We refined a previous estimate for the magnetic field strength on the
plane of sky in this region. Interferometric observations of polarized
emission from dust from \citet{Cortes2006a} estimated a magnetic field
of $B_{\mathrm{pos}}$=1.7\,mG, by using the Chandrashekar-Fermi technique. The
estimation needs a reliable tracer for the gas turbulent motions from the
region traced by the dust emission. By using the line-width from our
ion optically thin molecule H$^{13}$CO$^{+}(J=4\rightarrow3)$, $\Delta V=3$ \kms,  
% and H$^{13}$CN$(J=4\rightarrow3)$
%observations, where both lines present close FWHM velocity dispersion of
we were able to obtained an improved
estimation of $B_{\mathrm{pos}}=$855\,$\mu$G for the magnetic field,
which also refined the mass-to-magnetic flux ratio to $\lambda$=1.9 or
super-critical.  Along with $\lambda$, we calculated the contribution
from turbulence to the virial mass, getting a value of
M$_{\mathrm{submm}}$/M$_{\mathrm{vir}}=5.9$.  These two results suggest
that neither the magnetic field nor the turbulence have enough energy to
counterbalance gravity. Therefore, the G30.79 FIR 10 clump must be bound and
undergoing gravitational collapse.  This also reinforce the core as an
infall candidate.  
%Additionally, both ratios suggest that the magnetic
%field and the turbulence are in equipartition.

\begin{acknowledgements}
  P. C. Cort\'es and R. Parra acknowledges support by the FONDECYT
  grants 3085039 and 3085032 respectively. J. R. Cort\'es  and E. Hardy acknowledge
  support from the National Radio Astronomy Observatory of the United
  States.
\end{acknowledgements}

%\bibpunct{(}{)}{;}{a}{}{,} %
\bibliographystyle{aa}
\bibliography{biblio}

\end{document}